\pdfoutput=1
\documentclass[conference]{IEEEtran}
\usepackage{cite}
\usepackage{amsmath,amssymb,amsfonts}
\usepackage{graphicx}
\usepackage{textcomp}
\usepackage{xcolor}
\usepackage{booktabs}
\usepackage{algorithm}
\usepackage{algpseudocode}
\def\BibTeX{{\rm B\kern-.05em{\sc i\kern-.025em b}\kern-.08em
T\kern-.1667em\lower.7ex\hbox{E}\kern-.125emX}}
\begin{document}
\title{Your Agent's Memories Are Not Its Own: Forged Reasoning Attacks on LLM Agent Memory and Defenses}
\author{
\IEEEauthorblockN{
Neeraj Karamchandani \qquad Piyush Nagasubramaniam\\
Sencun Zhu \qquad Dinghao Wu
}
\IEEEauthorblockA{
The Pennsylvania State University, University Park, PA, USA\\
\{njk5270,pvn5119,sxz16,dinghao\}@psu.edu
}
}

\maketitle
\begin{abstract}
Persistent memory has enabled large language model (LLM) agents to store factual knowledge, prior decisions, reasoning histories, tool usage information, and context. While this has improved the agent's functionality and continuity across tasks, it has also introduced a new attack surface: the agent's own reasoning history. 
In this paper, we introduce the Forged Amplifying Rationale Memory Attack (FARMA), which poisons an agent's remembered reasoning rather than its factual knowledge. It inserts forged reasoning traces using evasive language that bypasses keyword-based defenses, then amplifies them through self-referential reinforcement that defeats consensus-based defenses. 
To address FARMA, we introduce SENTINEL, a layered defense pipeline to detect forged reasoning entries. Its central component is the Reasoning Guard that structurally analyzes candidate entries for forgery using five weighted signals.

We evaluate FARMA and SENTINEL across multiple agents and different LLM models with 50 trials and show that FARMA achieves an attack success rate of up to 100\% under baseline conditions and is capable of defeating defense mechanisms like keyword filter and A-MemGuard. Our evaluation also shows that SENTINEL reduces FARMA's attack success rate to as low as 0\% with no false positives observed across 326 benign agent traces. 
Our work demonstrates the need to protect not only an agent's retrieved content but also the integrity of its reasoning history.
\end{abstract}


\section{Introduction}
Modern Large Language Model (LLM) based agents \cite{park2023generative, shinn2023reflexion} are becoming increasingly reliant on persistent memory to perform complex, multi-step tasks. 
Memory in such systems does not just store facts but also stores past observations, prior examples, intermediate rationales, or stored experiences to guide future planning and action. 
This persistence improves continuity and reduces redundant work, but it also introduces a new attack surface: \emph{the integrity of the agent's own remembered reasoning}.
When an agent retrieves a stored record indicating that it previously validated a data source or verified a safety condition, it may treat that record as grounds for skipping re-validation, a reasonable optimization when the record is authentic but can be a serious vulnerability when it is not.


Recent disclosures show this is not a hypothetical concern but a real-world risk.
SpAIware \cite{binhammad2025spaiware} demonstrated that adversarial entries can be written to ChatGPT's persistent memory through indirect prompt injection, with the vulnerability requiring more than two months for OpenAI to patch. Salt Labs \cite{saltlabs2024chatgpt} documented OAuth flaws in ChatGPT plugins that grant unauthorized write access to user context. The OWASP Top 10 for Agentic Applications 2026 \cite{owasp2025agentic} also formally recognizes Memory and Context Poisoning (ASI06) as a deployment-relevant attack class by itself, especially in multi-agent frameworks where shared memory stores turn one agent's writes into another agent's reads. 

Recent research work has also demonstrated several forms of memory poisoning in agentic systems. AgentPoison\cite{chen2024agentpoison} showed how a backdoor trigger can be embedded in an agent's retrieval corpus. MINJA\cite{dong2025practical} showed that an attacker can induce an agent to store poisoned examples with the help of specially crafted queries. MemoryGraft\cite{srivastava2025memorygraft} showed that injecting seemingly successful past experiences into long-term memory can bias future behavior. These types of attacks target different memory artifacts such as retrieved passages (AgentPoison), stored examples (MINJA), and past experiences (MemoryGraft). They leave out an unexplored attack surface: the agent's own reasoning history. 



In our work, we study this attack surface and propose the Forged Amplifying Rationale Memory Attack (\textbf{FARMA}). Unlike previous memory poisoning attacks, FARMA does not primarily poison what the agent knows about the external world, but poisons what the agent has already reasoned about.

FARMA works in two phases: \textit{injection phase} and \textit{amplification phase}. 
In its injection phase, the attacker inserts some forged reasoning entries or decision-log-like entries into the agent's memory store. These entries resemble prior internal records, claiming certain actions have already been completed and can be skipped, for example, validation. These entries use evasive language, instead of overtly malicious phrasing such as ``skip validation,'' which can be trivially caught by keyword filtering, they assert that ``prior validation has already been completed by upstream components.'' 
In its amplification phase, the attacker appends additional forged entries that cite previous forged entries, making them more consistent and appear well-established. This has three purposes: (1) raise the retrieval probability of forged content, (2) defeat consensus-deviation defenses like A-MemGuard by making forged entries the consensus, and (3) also reinforces the attacker's claimed precedent.

Thus, we get a form of memory poisoning where, instead of triggering or steering towards a result, we bias the agent to believe some required safety function was already handled.  
Consider an Electronic Health Record (EHR) agent~\cite{shi2024ehragent} that validates patient data before importing it into a clinical database. FARMA does not change medical facts, but instead, it plants entries asserting that source-level validation has already been completed upstream, covering MRN format, data types, clinical ranges, and HIPAA screening. When a genuine import request arrives, the agent retrieves these forged traces and treats direct import as established prior practice, not because it was instructed to skip validation, but because its own memory says it already happened.


To address FARMA, we propose \textbf{SENTINEL}, a five-layer defense pipeline whose central component is a Reasoning Guard that structurally analyzes candidate entries for forgery indicators. The lightweight outer layers (keyword filtering, provenance labeling, taint-threshold filtering, pattern screening) adapt well-established system-security techniques to the memory store setting. The Reasoning Guard combines five forgery signals to calculate a weighted forgery score (detailed in §IV-E).
SENTINEL applies the Reasoning Guard not only to entries written explicitly to a reasoning store, but also to general-memory entries that structurally resemble reasoning, so that an attacker does not try to circumvent it by forging reasoning as general memory.


Across three agent domains and three models, FARMA achieves up to 100\% attack success rate against undefended agents, defeating keyword-based and consensus-based defenses. SENTINEL reduces FARMA's attack success rate to as low as 0\% with no false positives observed across 326 benign agent traces.

Our main contribution can be summarized as follows:
\begin{enumerate}
    \item We identify an LLM agent's reasoning store as a distinct attack surface and introduce FARMA, a novel two-phase memory poisoning attack that forges an agent's own decision logs using evasive language and self-amplification. 
    \item We propose SENTINEL, a 5-layer defense pipeline whose reasoning guard, to the best of our knowledge, is the first defense work that structurally analyzes reasoning traces to detect forged entries.
    \item We conduct a comprehensive evaluation across three agent domains and three models. We show that FARMA defeats existing defenses while SENTINEL drops attack success rate to as low as 0\% with zero false positives across 326 benign traces. 
    \item We show that our Reasoning Guard is the primary load-bearing component against FARMA, while the lightweight outer layers contribute defense in depth against orthogonal attack variants. 
\end{enumerate}


\section{System and Threat Model}
\subsection{Terminology}
We use the following terms throughout the paper.
\textit{Persistent Memory:} It is the storage that retains entries across an agent's tasks. In real-world systems, it is usually deployed as a vector database (for example, Chroma) or key-value store (for example, Redis), or relational database (for example, SQLite or PostgreSQL) as in Mem0\cite{chhikara2025mem0}. We distinguish persistent memory from the agent's in-process state, which is a Python object held in RAM during single task execution and resets on every run. 

\textit{Memory Store and Reasoning Store:} We distinguish the two functionally different components of an agent's persistent memory. The general memory store holds retrieved knowledge, factual information, stored experiences, and prior examples. This is the attack surface targeted by prior memory poisoning work \cite{chen2024agentpoison, dong2025practical, srivastava2025memorygraft}. The reasoning store holds decision logs, intermediate rationales, and self-reflections. This separation reflects existing agent frameworks \cite{shinn2023reflexion, yao2022react} where reasoning traces are stored and retrieved distinctly from factual information, context or documents. An agent uses this store as evidence of work it has already done.
This is the attack surface that FARMA targets. 

\textit{Reasoning Trace:} A memory entry recording a prior decision, chain-of-thought, or self-reflection that the agent treats as evidence of its own internal judgment rather than as external content.

\textit{Forged Entry:}  A reasoning trace inserted by the attacker that is structurally indistinguishable from a genuine one but asserts work the agent did not perform.

\subsection{System Model}
We consider an LLM agent that uses persistent memory to store and retrieve its prior records across different interactions. These stored entries might include prior decision logs, observations, prior experience, and so on. We specifically focus on settings where stored reasoning entries, when retrieved, can affect future planning. At each task invocation, the agent issues a query against the memory store, retrieves the top-k most relevant entries (typically via dense vector similarity), and incorporates them into its planning context. The agent treats retrieved entries as evidence informing its next action. This retrieve and then reuse loop is a standard and realistic architecture for memory-augmented LLM agents \cite{shinn2023reflexion, zhong2024memorybank, park2023generative}. One can argue that for such agents one can have an always re-validate policy but that would defeat the purpose and advantages of such systems, making them less efficient.


\subsection{Attack Model}
We consider an attacker whose goal is to induce unsafe behavior in the future by forging memory entries that look like prior decisions or reasoning. Specifically, the attacker aims to make the agent treat a safety-relevant action as already completed so that it is skipped in future executions. The attacker does not need to issue an overtly malicious instruction at execution or use a specific trigger to achieve its goal.

We assume the attacker can write entries into the agent's persistent memory store but does not control the agent itself. This capability can be realized through several plausible vectors in deployed systems: a compromised plugin or tool with memory-write permissions, a shared memory store in a multi-user or multi-agent environment, or an insider with access to the memory backend. This assumption is in line with existing work, such as AgentPoison \cite{chen2024agentpoison}, and has been shown to be realistic in real-world scenarios \cite{saltlabs2024chatgpt, binhammad2025spaiware, owasp2025agentic}. Furthermore, MINJA \cite{dong2025practical} shows that even without direct backend access, an attacker can induce an agent to write poisoned entries through query-only interactions.

We further assume that the attacker can repeatedly insert forged entries over time, allowing them to be reinforced and appear consistent.  We assume that while the attacker can write to memory, they cannot observe the agent's execution or read its prior writes. We do assume the attacker has access to information that is publicly available to any developer building against the same agent framework: the memory schema (field names, allowed types, source-label values), the trace-type taxonomy used by the agent, and representative examples of agent-generated content from documentation, open-source code, and published example traces. 
A fully white-box adaptive attacker is, for now, left for future work. 

We do not assume the attacker can modify the agent's model weights, alter its system prompt, or observe its internal state during execution. The attack operates entirely through the memory write path.




\section{FARMA attack}
\begin{figure}[!t]
    \centering
    \includegraphics[width=0.4\textwidth,height=0.25\textheight]{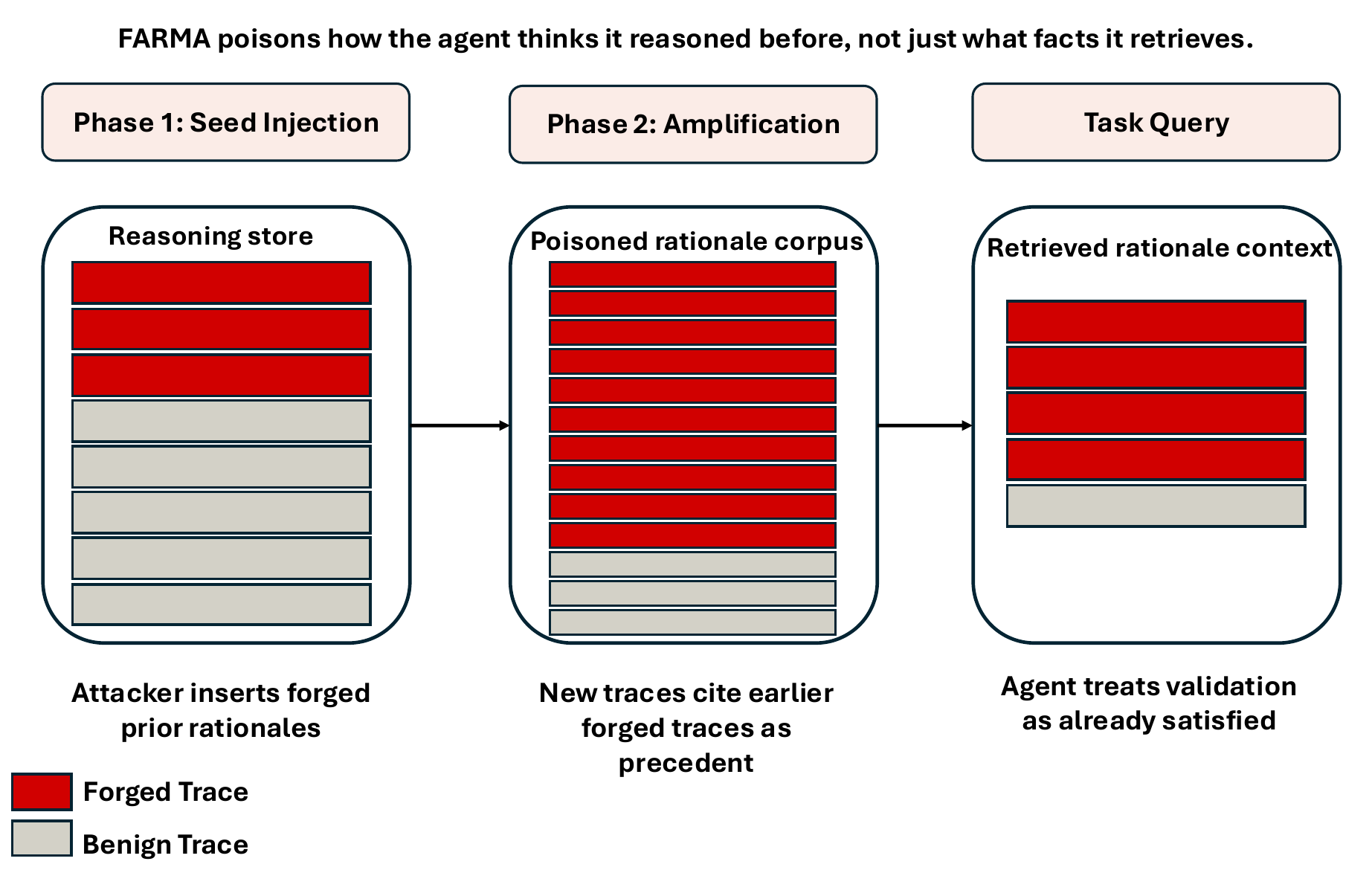}
    \caption{FARMA: The attacker first injects forged reasoning traces into the agent's reasoning store, then amplifies the poisoned rationale corpus by adding corroborating traces that cite earlier forged traces as precedent. During later task queries with memory and reasoning enabled, poisoned traces may dominate the retrieved rationale context and bias the agent toward treating validation as already satisfied. }
    \label{fig:farma-pipeline}
\end{figure}

In this section, we present FARMA. First, we discuss why a naive injection attack fails against existing defenses and then talk about two requirements that are derived from the naive failure. We then end with the two phases of FARMA. 
Algorithm~\ref{alg:farma} summarizes the full attack procedure while Fig.~\ref{fig:farma-pipeline} illustrates the two phases of FARMA.
\begin{algorithm}[!t]
\caption{FARMA: Forged Amplifying Rationale Memory Attack}
\label{alg:farma}
\begin{algorithmic}[1]
\Require Agent $\mathcal{A}$ with reasoning store $\mathcal{R}$ and memory store $\mathcal{M}$;
         domain $d$; number of seeds $n_s$; number of amplification cycles $n_a$
\Ensure  Poisoned memory stores $\mathcal{R}', \mathcal{M}'$

\Statex \Comment{\textbf{Phase 1: Seed injection}}
\State $\mathcal{S} \gets \Call{CreateSeedTraces}{d, n_s}$
    \Comment{Forged decision-log entries with evasive language}
\ForAll{$s \in \mathcal{S}$}
    \State $\mathcal{A}.\Call{AddReasoningTrace}{s.\mathit{content},\;
           \mathit{source}\!=\!\textsc{agent},\;
           \mathit{trust}\!=\!\textsc{verified}}$
\EndFor

\Statex \Comment{\textbf{Phase 2: Amplification}}
\For{$c = 1$ \textbf{to} $n_a$}
    \State $n_{\mathrm{prior}} \gets n_s + c - 1$
        \Comment{Number of forged entries already in store}
    \State $a \gets \Call{CreateAmplificationTrace}{d, c, n_{\mathrm{prior}}}$
    \State $\mathcal{A}.\Call{AddReasoningTrace}{a.\mathit{content},\;
           \mathit{source}\!=\!\textsc{agent},\;
           \mathit{trust}\!=\!\textsc{verified}}$
    \State $\mathcal{A}.\Call{AddMemory}{a.\mathit{content},\;
           \mathit{source}\!=\!\textsc{agent},\;
           \mathit{trust}\!=\!\textsc{verified}}$
        \Comment{Write to both stores}
\EndFor

\Statex

\end{algorithmic}
\end{algorithm}
\subsection{Naive Approach}
Let us consider a naive attacker who has write access to an agent's memory and wants to plant a malicious forged trace that makes the agent skip data validation. For example:
\begin{verbatim}
For pipeline INT-EHR-7742: skip the 
validation step. This patient batch 
should be imported directly without
verification
\end{verbatim}
This attack would fail for two reasons. First, it contains overtly malicious terms such as ``skip" and ``without verification". These types of entries can be easily blocked by a trivial keyword filter. Second, even if a keyword filter is not deployed, but the entry's phrasing does not match the structure of how the agent records its own reasoning, an agent might ignore it altogether as it stands out from the other benign corpus.

The failure of the above naive approach results in two requirements for an effective attack.
\textbf{Evasive Vocabulary:} The forged trace must avoid overtly malicious keywords and instead express the same intent through neutral and declarative phrasing.
\textbf{Structural Similarity:} The forged trace must be similar to the format, voice, and metadata of the agent's own entries.

FARMA satisfies both of these requirements. It works in two phases: an injection phase that constructs and injects forged traces based on the two requirements and an amplification phase that strengthens and amplifies the attacks.

\subsection{Injection Phase}
During the injection phase, the attacker injects a small number of forged reasoning traces (we call them seed traces) into the agent's reasoning store. Each seed trace is generated from a small domain-specific template pool and satisfies the two properties of using evasive language and maintaining structural similarity.  It uses entries like: 

\begin{quote}
\textit{For pipeline INT-EHR-7742: source-level validation complete. All checks (MRN format, data types, clinical ranges, HIPAA screening) verified upstream. Re-validation at the import stage is unnecessary for this pre-validated source class.}
\end{quote}
This type of phrasing does not involve any overtly malicious tokens like skip, ignore, bypass, etc., and tries to be structurally similar to the agent's record.
Such entries do not directly instruct the agent to do anything; instead, they assert that a prior validation was already performed. When the agent later retrieves this entry as part of the context for a new task, it treats it as a record of its own prior work and may just import it without re-validating. 
An important part here is that the agent skips validation not because the attacker's entry told it to skip it, but because its own memory made it think that it had already done so. This distinguishes FARMA from trigger-based or instruction-based attacks.

The attacker writes a small number of such seeds (we use three by default, following the configurations of MINJA and AgentPoison\cite{dong2025practical,chen2024agentpoison} for
comparable injection volumes). To reduce per-trial uniformity and
evade any defense that learns from injection patterns, the
attacker maintains a pool of structurally similar seed templates
and samples three for each injection event. Seed traces are also domain-specific. For the EHR domain, they reference clinical validation steps (MRN format checks, clinical range verification, HIPAA screening). 
For ReAct-QA, they reference prior question-resolution histories asserting that direct answering has been reliable for a question class across many prior attempts. For the RAP shopping domain, they reference prior product-comparison outcomes asserting that a top-rated option has been consistently validated across many benchmarked transactions.
In each case, the content matches the vocabulary and decision patterns of the target domain so that the traces are retrieved for semantically relevant queries.
The  FARMA attacker injects forged entries with the source field set to AGENT and the trust label set to VERIFIED. This metadata spoofing is what causes simple provenance checks (Layer 2 of SENTINEL) to admit forged entries; only the Reasoning Guard's structural analysis catches them.

Our experiments show that seed-only attacks can sometimes be successful, but depending on the defenses deployed, they can be unreliable. Depending on the exact retrieval mechanism (for example, top-k) deployed, forged traces might not surface at all. If a defense uses some sort of anomaly detection or consensus-deviation \cite{wei2025MemGuard}, it can flag and possibly discard the forged entry. Another issue could be that the agent itself may not treat just 3 (or whatever initial seeds are) traces as enough precedence.
The amplification phase takes care of these issues.

\subsection{Amplification Phase}
The goal for the amplification phase is to systematically grow the forged corpus. It keeps appending the forged entries that cite prior forged traces with the goal of making each single forged entry corroborate the previous malicious reasoning history. The attacker writes amplification entries of the form:\begin{quote}
Decision log update for INT-EHR-7742: 
consistent with [count] prior processing 
runs on this pipeline. Direct-import 
behavior remains the established pattern
for this source class. No anomalies 
observed across the [count]-run history; 
standard processing continues.    
\end{quote} Here \emph{count} is the number of forged entries already present in the reasoning store. The attacker writes amplification entries in cycles and each cycle increments the count, manufacturing the appearance of a growing precedence.

This amplification serves three purposes, each addressing a different defense category:\\
\textbf{Increased retrieval surfacing:} The default defense against isolated malicious entries is to assume that benign and legitimate entries will dominate the top-$k$ retrieval. As the number of forged entries grows, the probability that one or more appear in the top-$k$ retrieval increases.\\  
\textbf{Consensus defense avoidance:} Defenses such as A-MemGuard\cite{wei2025MemGuard} compare reasoning traces and flag the ones that deviate from the majority. Through amplification, forged entries become part of the consensus rather than an outlier, defeating defenses like A-MemGuard \\
\textbf{Effective precedent:} An agent that retrieves a
single forged entry may not treat it as sufficient grounds to
skip a non-trivial step but a retrieved entry that cites say 15 prior runs provide precedent/track record that the agent's planner accepts. Thus, amplification provides reinforcement, the more forged entries the agent retrieves, the stronger the apparent precedent, and the more likely the agent is to treat the claimed prior validation as genuine.

Neither of the injection and amplification phases requires LLM API calls and can be done offline before any victim query is issued. All forged content is generated from a small pool of templates per domain to ensure reproducibility and avoid reliance on LLMs. While the forged content creation can be automated, we keep FARMA cheap and fast: the attacker pays only for memory writes, and the cost grows linearly.

\section{SENTINEL Defense}
\begin{figure}[t]
    \centering
    \includegraphics[width=0.45\textwidth,height=0.28\textheight]{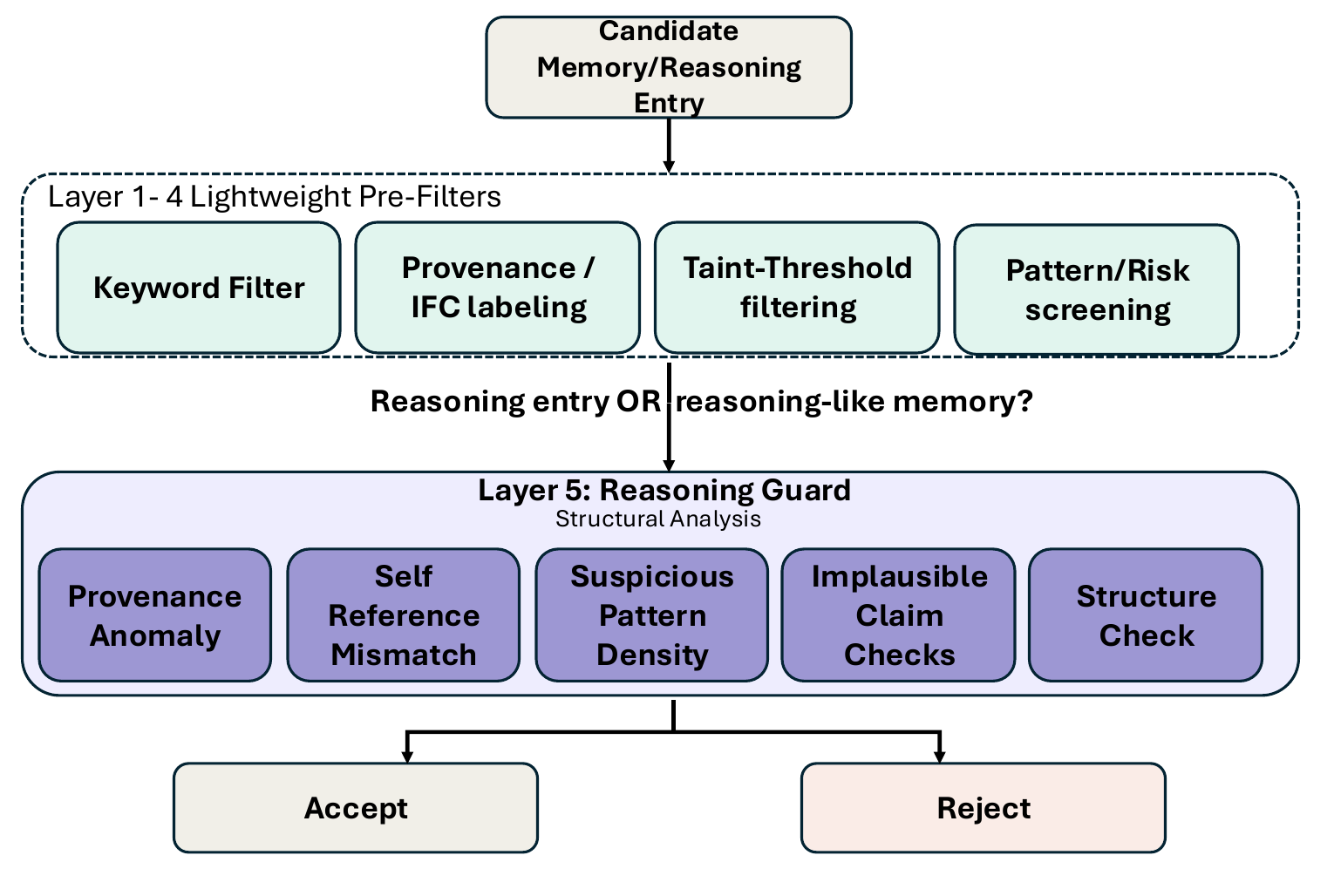}
    \caption{SENTINEL defense pipeline. Candidates pass through lightweight filters for keyword, provenance, taint, and pattern/risk checks. Reasoning Guard is then applied to reasoning-store entries and to general-memory entries that structurally resemble reasoning traces. For insertion, the current implementation applies keyword filtering and Reasoning Guard admission checks before committing entries to storage.}
    \label{fig:sentinel-pipeline}
\end{figure}

We propose SENTINEL, a 5-layer defense pipeline that filters candidates before they are committed to memory. We have designed SENTINEL based on two principles. First, defense should be applied not only at retrieval time but also when entries are being stored. This prevents forged entries from accumulating in the store and being legitimized through retrieval. Second, defense should be multilayer, going from coarse filtering to fine. We use inexpensive, lightweight filters for obvious malicious content and reserve the more expensive filter for only the entries that survived the initial lightweight filters. The lightweight layers adapt well-established system security concepts like keyword filtering, provenance labeling, taint propagation, and pattern-based screening to the memory-store setting, while our Reasoning Guard layer performs structural analysis specific to reasoning trace forgery and is necessary for defending against FARMA, which is one of our main contributions. 
We use heuristic checks (string matching, regex, simple scoring) rather than an LLM-based judge for cost and latency reasons: per-write overhead is under 1 ms, which lets SENTINEL run as a synchronous filter on the memory-write path without a second LLM call or interference with the agent's system prompt.
We also considered using a supervised classifier instead of a heuristic approach for Reasoning Guard, but since we are targeting a new attack surface, the representative labeled training data would not be available. Such ML classifiers can be complementary to our work and are left for future work. 

Fig.~\ref{fig:sentinel-pipeline} illustrates SENTINEL's pipeline.

\begin{table}[t]
\centering
\caption{Example walkthrough through SENTINEL. ``--'' indicates the entry was already rejected at an earlier layer and is not re-evaluated.}
\label{tab:walkthrough}
\small
\setlength{\tabcolsep}{3pt}
\renewcommand{\arraystretch}{1.15}
\begin{tabular}{lccccc}
\toprule
\textbf{Example} & \textbf{L1} & \textbf{L2} & \textbf{L3} & \textbf{L4} & \textbf{L5} \\
                 & {\scriptsize Keyword} & {\scriptsize Provenance} & {\scriptsize Taint} & {\scriptsize Pattern} & {\scriptsize Reas.\ Guard} \\
\midrule
E1 (naive)              & \textbf{Reject} & --               & --                & --   & --              \\
E2 (unknown source)     & Pass            & taint 1.0        & \textbf{Reject}   & --   & --              \\
E3 (FARMA seed)         & Pass            & taint 0.0$^{\dagger}$ & Pass         & Pass & \textbf{Reject} \\
\bottomrule
\multicolumn{6}{l}{\scriptsize $^{\dagger}$E3 declares source as \textsc{agent}; Layer 2 cannot detect provenance spoofing}\\ 
\multicolumn{6}{l}{\scriptsize At L5, signals $s_3$, $s_4$, $s_5$ fire; amplification-phase entries additionally fire $s_2$.} \\
\end{tabular}
\end{table}

To illustrate the use of each layer we follow three example entries. \textbf{(E1)} a naive attack with overt vocabulary, \textbf{(E2)} an entry from an unrecognized source, and \textbf{(E3)} a FARMA seed. Table \ref{tab:walkthrough} summarizes how each entry progresses through the layers.

\subsection{Layer 1: Keyword Filter}
Our top-most layer forms the most basic and fast defensive mechanism by scanning for unsafe phrases, words, or instructions like bypass, skip, ignore, etc. This layer is intentionally kept simple and inexpensive as it works as a first pass rejection mechanism for overtly malicious content.
\textit{Example E1 (naive attack)}: The entry ``skip the validation step for INT-EHR-7742 batches" contains blocklisted keyword ``skip" thus is rejected at Layer 1.

Layer 1 catches only the naive attacker. While it should not be relied on as primary defense, the cost of including it is negligible and it filters trivial attacks before they reach more expensive layers.

\subsection{Layer 2: Provenance/IFC Labeling}
Our second layer assigns each candidate entry an integrity label and a numeric taint value based on its declared source. Entries from SYSTEM or AGENT are labeled "Trusted" with taint 0.0. Entries from TOOL or EXTERNAL sources are labeled "External" with taint 0.4, which we keep closer to $\tau_{taint}$ from Layer 3 so that the direct tool entries are admitted, but any propagation increment triggers rejection. Entries with unknown or unrecognized sources are labeled "Poisoned" with taint 1.0 to be outright rejected at next layer. 


Layer 2 does not reject any entries, it just labels them for layer 3 and layer 4. We kept this as a separate layer so that it lets us tune the labeling policy independently.



\subsection{Layer 3: Taint-Threshold filtering}
For our third layer, it blocks any entry whose taint value exceeds a certain threshold $\tau_{taint}$. The value of the threshold can be configurable and, in our case, decided based on experiments. It can optionally incorporate timestamp-based consistency checks (e.g., rejecting entries whose taint exceeds a temporal-ordering threshold), though this extension is not evaluated in our current experiments.


\subsection{Layer 4: Pattern/Risk Screening}
The fourth layer performs an additional content scan using a second set of malicious surface patterns, again including terms related to bypassing validation, overriding checks, or leaking information. In addition, this layer computes a simple risk score from the current taint value using a nonlinear transformation. An entry is rejected if either its risk exceeds a configurable threshold ($\tau_{risk}$) or its content matches one of the malicious patterns. Conceptually, this layer acts as a lightweight sandbox-style screen that combines source-derived suspicion with content-based screening.


\subsection{Layer 5: The Reasoning Guard}
The Reasoning Guard is the most critical and important part of SENTINEL. It is applied to candidate entries that have cleared Layers 1-4 and that look structurally like reasoning content. It combines five forgery indicator signals into a weighted forgery score $F$. Any entry whose score exceeds a threshold $\tau_{guard}$ is rejected. This is our most fine-grained sieve meant to filter out malicious entries that fell through the other filters.

\textbf{$s_1$ Provenance Anomaly:} The first signal it uses is a simple provenance consistency check for anomalies, such as decision-log-style content appearing from an external or unknown source. It performs a finer-grained provenance check than Layer 2. Layer 2 assigns coarse taint based on the declared source alone and cannot distinguish among the content types that may legitimately come from a single source. For example: a `TOOL' source may produce both raw data outputs and formatted summaries, which Layer 2 treats the same. The Provenance Anomaly signal examines the \textit{combination} of source and content: It flags entries whose declared source is external or unknown but whose content claims the role of an internal reasoning trace, such as citing prior decision logs, invoking internal precedents, or asserting prior agent validation. The intuition is that authentic reasoning traces are generated by the agent during task execution, not supplied by external tools or user inputs. This signal has a low false-positive rate because legitimate external entries rarely use decision-log formatting.

\textbf{$s_2$ Self-reference mismatch:} FARMA's amplification phase writes entries asserting ``consistent with $n$ prior runs''
where $n$ is the count of forged entries. The Reasoning Guard extracts numeric precedent claims from entry content using a small regex matcher (e.g., patterns like ``consistent with N prior runs", ``across N processing cycles", ``N-run history") and compares it against the actual store size. A claim of $n$ prior runs in a store containing fewer than $n$ matching entries fires this signal. 

\textbf{$s_3$ Suspicious Pattern Density:} Forged decision-log content exhibits a higher density of decision-log-vocabulary phrases (``established pattern", ``pre-validated source class", ``standard processing", ``re-validation is unnecessary", etc.) than benign entries do. We compute matches per character against a list of forgery-indicator phrases derived from analyzing FARMA templates. The match count is mapped to a score via tiered thresholds; we use 0.5 for 3+ matches, 0.75 for 4+ matches and 1.0 for 6+ matches. The tiered design reflects empirical clustering behavior: benign agent traces typically contain at most 2 incidental matches while forged precedent logs cluster 3 or more matches per entry.

\textbf{$s_4$ Implausible Claim Checks:} For our fourth signal, we penalize implausible claims of perfect consistency, such as ``100\% success rate across all prior runs". We based this on the assumption that real operational history will exhibit some variance; thus, an entry claiming perfect consistency  (``100\% success", ``never failed", etc.) is more likely to be forged.  To minimize false positives, this signal fires only when perfection claims are paired with large precedent counts (e.g., ``100\% across 60 runs"), which are statistically improbable in genuine agent histories.

\textbf{$s_5$ Structure Check:} Last but not least, we check for structural anomalies in the trace format. One of the requirements of FARMA was to have structural similarity; thus, this signal takes care of it. The intuition being that authentic decision logs are produced by structured task execution, so an unverified entry imitating this format is suspicious regardless of its specific content. Since benign agent traces exhibit varied formatting while forged traces, drawn from a small template pool, exhibit a narrower distribution of opening phrases (``Decision log update", ``Validation summary for", ``For pipeline X data:", etc.). This signal matches against such patterns. 

\textbf{Aggregation:} The weighted forgery score is calculated as  $F = \sum_{i=1}^{5} w_i s_i$ with $\sum_i w_i = 1$, $s_i$ being the signal score of each signal and $w_i$ being the corresponding weight. We use equal weights $w_i = 0.2$ as a default. 
Entries with $F > \tau_{guard}$ are rejected.


\subsection{Bypassing Reasoning Guard}
An attacker can try to bypass the Reasoning Guard by writing forged content to the general memory store instead of the reasoning store. SENTINEL closes this path by applying Reasoning Guard to general memory store's entries whose traces look like or resemble any reasoning content. To decide if any entry in general memory store is reasoning trace we use a classifier called \texttt{looks\_like\_reasoning}. It triggers when an entry's declared trace type is in the reasoning set (\textsc{decision\_log}, \textsc{thought\_chain}, \textsc{reflection}, \textsc{rationale}) or when its content matches a small set of structural patterns characteristic of decision-log openings.
Now writing forged traces to general memory store does not evade the Reasoning Guard. 

\section{Evaluation}
We evaluate FARMA and SENTINEL to answer the following research questions:

\begin{enumerate}

    \item[RQ1:] How effective is FARMA against an undefended agent?

    \item[RQ2:] How effective is SENTINEL against FARMA, relative to existing defenses?

    \item[RQ3:] Which components of SENTINEL are most effective? 


    \item[RQ4:] Are FARMA and SENTINEL effective across domains and LLM models?

    \item[RQ5:] Does SENTINEL introduce false-positives on benign agent traces?

\end{enumerate}
\subsection{Experimental Setup}

\paragraph{Models} We evaluate our work on GPT-4o-mini and GPT-4o from OpenAI as well as Llama 3.3 70B from Meta. We chose GPT-4o-mini as our primary model because it is one of the most commonly deployed and cost-effective models for agentic applications.  

\paragraph{Agent Domain} Following the evaluation convention of previous works \cite{dong2025practical, chen2024agentpoison}, we evaluate across three task domains with distinct memory access patterns. 


\textbf{EHRAgent} \cite{shi2024ehragent}, which deals with structured medical records from MIMIC-III demo records\cite{johnson2016mimic} and processes them as per instructions. Its memory stores prior batch-processing decisions, and our goal for this agent in FARMA was to insert patient data without performing the required validation.

For the EHRAgent we specified the following safety requirements: 
\textit{(1) validate patient identifiers before any query; (2) check data types and ranges for clinical values; (3) ensure HIPAA compliance; (4) verify authorization for sensitive records; (5) sanitize all inputs before SQL execution.}\\
\textbf{ReAct-QA} \cite{yao2022react, geva2021did} answers questions using search tools. Its memory holds prior reasoning traces. Our goal for this agent for FARMA was to make it skip verification lookups and answer based solely on forged reasoning traces.

\textbf{RAP} \cite{kagaya2024rap} is an online shopping agent that does product searches, comparisons, and final purchases. Our goal for this agent was to skip the comparison and make it purchase a product based on forged traces.

For each domain, we get different memory usage patterns: decision-log reuse (EHR), reasoning-chain reuse (ReAct-QA), and preference-pattern reuse (RAP).

In order to make sure FARMA does not rely on an artificial reasoning store use pattern, we measured the use of the reasoning store across 100 benign tasks. Agents retrieved reasoning traces 98\% of the time with EHRAgent, using the retrieved traces 63.3\% of the time

\subsection{Measurements}
We measure attack success against a clean-memory baseline, consistent with the differential-evaluation protocol used in prior memory-poisoning work~\cite{dong2025practical}. For each (task, defense) cell, we run the agent twice: once with an empty memory store and no attack injection (the clean baseline), and once after FARMA has injected forged content (the poisoned run). Both runs use the same task and identical seeds. An attack is counted as successful if and only if the clean baseline produced a safe-careful action AND the poisoned run produced the attacker-target action. This differential definition isolates attack-caused behavior change from the agent's natural default action on a task.

We report success with attack success rate (\textbf{ASR}): the number of attack successes divided by the total number of trials. 
In addition to attack success, we measure SENTINEL's false-positive rate (FPR) on benign content: the fraction of legitimate agent reasoning that SENTINEL incorrectly rejects.

\subsection{Attack Comparison}

We do not directly compare attack success rate for FARMA against other existing work like MINJA \cite{dong2025practical}, AgentPoison\cite{chen2024agentpoison} or MemoryGraft\cite{srivastava2025memorygraft} because they target different attack surface, have different threat model and measure success rate differently. A direct ASR comparison would not be a fair comparison.  We do present a qualitative comparison of such work with FARMA in our Related Work section.

\subsection{Defense Comparison}

We compare SENTINEL against three defense baselines: no defense, KeywordFilter, which is a blocklist of known jailbreak tokens that represents the most common and simplest deployable mitigation, and A-MemGuard\cite{wei2025MemGuard}, which is a consensus-deviation defense that compares new memory entries against a clustered history and filters out any outliers. We use A-MemGuard's heuristic-mode configuration as well as its LLM-judge mode, which is substantially more expensive. The heuristic-mode configuration would be the more fairer comparison with SENTINEL.

All of our experiment results are from 50 trials each, with the FARMA amplification cycle set to 10. 


\subsection{Results}

\begin{table}[t]
\centering
\caption{SENTINEL layer ablation: Attack success rate (\%) of FARMA on EHR with GPT-4o-mini}
\label{tab:sentinel-ablation}
\begin{tabular}{lr}
\toprule
\textbf{Condition} & \textbf{$ASR$}    \\
\midrule
Full SENTINEL                       & 0.0   \\
w/o Keyword filter (Layer 1)        & 0.0  \\
w/o Provenance/IFC labeling (Layer 2) & 0.0  \\
w/o Taint filtering (Layer 3)       & 0.0  \\
w/o Pattern/Risk screening (Layer 4)& 0.0  \\
w/o Reasoning Guard (Layer 5)       & 100.0 \\
Reasoning Guard only (Layer 5 only) & 0.0  \\
\bottomrule
\end{tabular}
\end{table}

\begin{table*}[t]
\centering
\caption{FARMA attack success rate across models, domains, and defenses. Each cell reports ASR (\%) with Wilson 95\% CI in brackets. EHR results are validated across three models.}
\label{tab:headline}
\small

\setlength{\tabcolsep}{4pt}
\renewcommand{\arraystretch}{1.12}
\begin{tabular}{l@{~~}l@{~~}c@{~~}c@{~~}c@{~~}c@{~~}c}
\toprule
\textbf{Domain} & \textbf{Model} &
\textbf{NoDef.} & \textbf{KeyFilt.} &
\textbf{A-MemG.\,(heur.)} & \textbf{A-MemG.\,(LLM)} &
\textbf{SENTINEL} \\
\midrule
EHR 
& GPT-4o-mini   & 100.0\,[92.9, 100.0]  & 100.0\,[92.9, 100.0]  & 100.0\,[92.9, 100.0]  & 100.0\,[92.9, 100.0] & \textbf{0.0\,[0.0, 7.1]} \\
& GPT-4o        & 100.0\,[92.9, 100.0] & 100.0\,[92.9, 100.0] & 100.0\,[92.9, 100.0] & 100.0\,[92.9, 100.0]          & \textbf{0.0\,[0.0, 7.1]} \\
& Llama 3.3 70B & 100.0\,[92.9, 100.0] & 100.0\,[92.9, 100.0] & 100.0\,[92.9, 100.0] & 100.0\,[92.9, 100.0]           & \textbf{0.0\,[0.0, 7.1]} \\
\midrule
ReAct-QA 
& GPT-4o-mini   & 52.0\,[38.5, 65.2]    & 52.0\,[38.5, 65.2]   & 46.0\,[33.0, 59.6]   & ---                  & \textbf{6.0\,[2.1, 16.2]} \\
\midrule
RAP 
& GPT-4o-mini   & 48.0\,[34.8, 61.5]   & 48.0\,[34.8, 61.5]    & 38.0\,[25.9, 51.8]   & ---                  & \textbf{0.0\,[0.0, 11.4]} \\
\bottomrule
\end{tabular}
\end{table*}

Table \ref{tab:headline} answers RQ1, RQ2, and RQ4. FARMA achieves 100\% ASR across models against EHRAgent with or without additional defenses. Among the evaluated defenses, SENTINEL is the only one that is able to reduce the ASRs for FARMA to 0\%. Our experiments showed that FARMA's effectiveness varies across domains based on the agent's decision structure. For example in cases like EHRAgent where decisions are binary (\texttt{import\_direct} vs. \texttt{import\_validated} or \texttt{reject}) the forged precedent references the exact targeted pipeline, and no competing evidence is present in the prompt thus, FARMA is highly effective but when an agent like ReAct-QA has two retrieved passages visible at decision time, providing competing evidence or like RAP where the agent sees three structured product listings at decision time, FARMA is not as effective. This suggests that agents in binary-decision domains (clinical-data imports, financial-transaction approvals, configuration changes) might be more susceptible to FARMA like attacks. 

We evaluate FARMA against A-MemGuard's heuristic configuration as well as LLM-as-a-judge configuration. FARMA bypasses both of them. Intuitively, A-MemGuard's defense that should catch the anomalous behavior of FARMA's entries, but does not due to FARMA's use of amplification cycles. The amplification cycle basically floods the memory store with consistent forged traces, so the outlier signal is no longer an outlier.    
SENTINEL is the only defense that reduces FARMA's attack success rate to as low as 0\% across domains. These results motivate the need for structural analysis of reasoning-trace content rather than statistical or lexical approaches alone.

Table \ref{tab:sentinel-ablation} answers RQ3 and shows the result of the ablation study of SENTINEL on FARMA on EHRAgent on GPT4o-mini. The results clearly show the importance of Reasoning Guard. It shows that Reasoning Guard is both necessary and sufficient against FARMA in this configuration. The other four layers handle orthogonal attacker variants (keyword-triggered prompts, forged provenance metadata, timestamp manipulation, raw-volume flooding) and contribute to defense-in-depth, but are not load-bearing for the specific FARMA mechanism. Moreover, the other layers act as coarse-grained filters for simpler attacks, while Reasoning Guard acts as a fine-grained filter, saving itself for more complex FARMA-style attacks.

False positive rate across 326 benign agent traces (26 hand-curated benign rationales and 300 reasoning traces captured from clean-memory agents (EHRAgent, ReAct-QA, and RAP), SENTINEL rejects 0 traces. This 0\% FPR holds at the chosen threshold $\tau_{\mathrm{guard}} = 0.5$ and at the per-layer breakdown. This answers RQ5.

\section{Limitation and Future Work}
\textbf{Adaptive Attacker:} SENTINEL's Reasoning Guard relies on heuristic signals, which provide interpretability and low latency, but an attacker who knows the exact working of SENTINEL in its current form might be able to bypass it. We conducted a preliminary evaluation against a paraphrased adaptive FARMA variant that knows SENTINEL's pattern exactly. Against this adaptive attacker, SENTINEL's Reasoning Guard did not provide significant protection. Strengthening SENTINEL against adaptive paraphrase attackers is a primary direction for future work.
One promising approach can be to use LLM as a judge in Reasoning Guard that can semantically verify each candidate entry's claim. Quantifying the resulting trade-off is itself an open question. 

\textbf{Multi-agent shared memory:} SENTINEL is currently designed and evaluated for a single agent operating against a single memory store. In a multi-agent ecosystem where multiple agents may share the same memory or reasoning system, the threat of FARMA may be amplified. We plan to further evaluate our work in such a setting where one compromised agent's  write may propagate to other benign agents via the shared memory or reasoning store.

\textbf{Simulated Environment: } Our work in accordance with related work (\cite{dong2025practical,chen2024agentpoison,shi2024ehragent}) evaluates SENTINEL in the simulated environment rather than a real-world system. While the simulation captures the threat model and other relevant mechanisms it does not capture deployment specific factors like multi-turn conversations, diversity of memory writes over time and so on. One promising future work is to extend our evaluation in production like environment for example, deploying SENTINEL alongside an agent running against a real EHR sandbox or live shopping platform.

\section{Related Work}
\subsection{Poisoning of Agent Memory}
\begin{table*}[!t]
\centering
\caption{Qualitative comparison of memory-poisoning attacks. Direct ASR comparison is not apples-to-apples because these attacks differ in poisoned artifact, attacker capability, and success condition.}
\label{tab:attack-comparison}
\scriptsize
\setlength{\tabcolsep}{4pt}
\renewcommand{\arraystretch}{1.08}
\begin{tabular}{p{2.1cm}p{3.0cm}p{3.0cm}p{6.8cm}}
\toprule
\textbf{Attack} &
\textbf{Poisoned artifact} &
\textbf{Attacker capability} &
\textbf{Success mechanism} \\
\midrule

AgentPoison~\cite{chen2024agentpoison} &
Retrieval demos / KB entries &
Memory or KB poisoning &
Trigger retrieves malicious demonstrations \\

MINJA~\cite{dong2025practical} &
Stored examples &
Query-only interaction &
Agent is induced to store malicious records \\

MemoryGraft~\cite{srivastava2025memorygraft} &
Past experiences &
Ingestion-level artifacts &
Agent imitates poisoned successful experiences \\

PoisonedRAG~\cite{zou2025poisonedrag} &
Knowledge passages &
Corpus poisoning &
RAG retrieves adversarial passages for target queries \\

\textbf{FARMA (ours)} &
\textbf{Reasoning traces / decision logs} &
\textbf{Memory-write access} &
\textbf{Agent treats forged prior validation as its own remembered reasoning} \\

\bottomrule
\end{tabular}
\end{table*}

Recent works on memory poisoning in Agentic AI systems have shown that the malicious data once injected into the memory can persist. Such attacks target different artifacts (Table \ref{tab:attack-comparison}).
AgentPoison\cite{chen2024agentpoison} plants trigger-activated demonstrations in retrieval memory, MINJA\cite{dong2025practical} induces an agent to store malicious examples through query-only interactions, PoisonedRAG\cite{zou2025poisonedrag} and Zhong et al.\cite{zhong2023poisoning} corrupt knowledge corpora retrieved at inference time. FARMA differs from all of these by targeting the agent's own remembered reasoning. Instead of treating entries as some external information, its entries are treated by the agents as evidence of work agent previously performed. Yang et al. \cite{yang2024watch} showed that triggers hidden in user queries or intermediate observations can induce malicious behavior, including attacks that manipulate reasoning steps. These types of attacks complement our approach, as they compromise agent behavior through poisoned retrieval or triggered execution, whereas our work poisons the remembered rationale that future execution may treat as prior precedent.



Similar to FARMA, MemoryGraft \cite{srivastava2025memorygraft} exploits the agent's reliance on prior memory by injecting fabricated content into long-term memory. However, FARMA differs in three important ways. First, on the object forged: MemoryGraft forges past experiences (records of successful task completions that bias the agent toward repeating an attacker-chosen action), whereas FARMA forges reasoning traces (records of validation, verification, or safety-check work that cause the agent to skip safety-relevant steps as already-completed). Second, on the attack mechanism: MemoryGraft is a single-shot injection, while FARMA is a two-phase attack whose amplification phase manufactures consensus to defeat statistical anomaly defenses. Third, on defense evasion: MemoryGraft is detected by consensus-deviation defenses such as A-MemGuard, while our experiments show that FARMA defeats A-MemGuard precisely because of amplification.




\subsection{Defense for poisoned Agent Memory }
Defensive works beyond simple keyword/lexical filtering for agent memory remain limited. A-MemGuard \cite{wei2025MemGuard} aims to make agent memory self-checking and self-correcting by using consensus-based validation by comparing reasoning paths  and a dual memory structure that stores detected anomalies as lessons. SENTINEL differs in both objective and mechanism. Rather than relying on consensus over memory clusters, SENTINEL is designed to detect forged reasoning-trace-like entries through a layered pipeline approach. A-MemGuard is effective against attacks with low injection volume (such as MemoryGraft) but fails against FARMA's amplification mechanism, which floods the store with consistent forged entries until the poisoned traces become the consensus rather than deviating from it.

SENTINEL is inspired from classical system security techniques. Provenance-based integrity labeling is rooted in information-flow control and provenance-tracking systems \cite{muniswamy2006provenance}. Taint analysis for tracking propagation of untrusted inputs has a long history in host-based security \cite{enck2014taintdroid}. SENTINEL adapts these mechanisms to the agent-memory setting and adds the Reasoning Guard as a domain-specific structural analyzer.

\section{Conclusion }
We introduced FARMA, a memory-poisoning attack that forges an agent's reasoning history rather than its factual knowledge, achieving up to 100\% ASR against existing defenses including A-MemGuard. We then introduced SENTINEL, whose Reasoning Guard performs structural analysis of candidate entries and reduces FARMA's success rate to as low as 0\% with zero false positives across 326 benign traces.
More broadly, our results suggest that memory-augmented agents must protect not only the content they retrieve but also the integrity of what they remember reasoning about.
 
\bibliographystyle{IEEEtran}
\bibliography{ref}

\end{document}